\documentclass[10pt,aps,prl,preprintnumbers,superscriptaddress,showpacs,twocolumn,floatfix]{revtex4-1}
\usepackage{graphicx}

\newcommand{\bfm}[1]{\mbox{\boldmath$#1$}}

\newcommand{\gsim}{\;\rlap{\lower 3.5 pt \hbox{$\mathchar \sim$}} \raise 1pt \hbox {$>$}\;}
\newcommand{\lsim}{\;\rlap{\lower 3.5 pt \hbox{$\mathchar \sim$}} \raise 1pt \hbox {$<$}\;}

\begin{document}

\title{
\boldmath Hyperfine splitting in positronium to ${\cal O}(\alpha^7m_e)$:
one-photon annihilation contribution
\unboldmath}
\author{M. Baker}
\affiliation{Department of Physics, University of Alberta, Edmonton, Alberta T6G 2J1, Canada}
\author{P. Marquard}
\affiliation{Deutsches Elektronen Synchrotron DESY, Platanenallee 6, 15738 Zeuthen, Germany}
\author{A.A. Penin}
\affiliation{Department of Physics, University of Alberta, Edmonton, Alberta T6G 2J1, Canada}
\affiliation{Institut f\"ur Theoretische Teilchenphysik,
Karlsruhe Institute of Technology, 76128 Karlsruhe, Germany}
\author{J. Piclum}
\affiliation{Physik Department,
Technische Universit\"at M\"unchen, 85748 Garching, Germany}
\affiliation{Institut f\"ur Theoretische Teilchenphysik und Kosmologie, RWTH Aachen, 52056 Aachen, Germany}
\author{M. Steinhauser}
\affiliation{Institut f\"ur Theoretische Teilchenphysik, Karlsruhe Institute of Technology,
76128 Karlsruhe, Germany}

\preprint{ALBERTA-THY-03-14,  DESY-14-011,  LPN14-045,
SFB/CPP-14-08, TTK-14-06, TTP14-005, TUM-HEP-930/14}

\begin{abstract}
We present the complete result for the  ${\cal O}(\alpha^7m_e)$ one-photon
annihilation contribution to the hyperfine splitting of the ground state
energy levels in positronium. Numerically it increases the prediction of quantum
electrodynamics by $217\pm 1$ kHz.
\end{abstract}

\maketitle
Positronium, an electromagnetic bound state of an electron  and a positron, is
the lightest known atom. The strong interaction effects in positronium are
suppressed by the small ratio of the  electron mass $m_e$ to the hadronic mass
scale, and the  properties of the bound state can be calculated perturbatively
in quantum electrodynamics (QED) as an expansion in Sommerfeld's fine-structure
constant $\alpha$, with very high precision only limited by the complexity of
the calculations. Positronium is thus a unique laboratory for testing the QED
theory of weakly bound systems. At the same time a deviation of the QED
predictions from the results of experimental measurements may be a signal of an
exotic ``new physics'' \cite{Glashow:1985ud}.

Positronium hyperfine splitting (HFS) defined by the mass difference between the
spin-triplet orthopositronium and spin-singlet parapositronium states, is among
the most accurately measured physical quantities. Already three decades ago HFS
in positronium has been determined with the precision of about ten parts in
a million  \cite{Mills:1983zzd,Ritter:1984} yielding
\begin{equation}
\Delta\nu^{\rm exp}=203.387\,5(16)\,\mbox{GHz}
\label{eq::exp1}
\end{equation}
and
\begin{equation}
\Delta\nu^{\rm exp}=203.389\,10(74)\,\mbox{GHz},
\label{eq::exp2}
\end{equation}
respectively.
Recently a new result with reduced  systematic uncertainty from the positronium
thermalization effect has been reported
\cite{Ishida:2013waa}
\begin{equation}
\Delta\nu^{\rm exp}=203.394\, 1(16)_{\rm stat.}(11)_{\rm syst.}\,\mbox{GHz},
\label{eq::exp3}
\end{equation}
which overshoots the previous measurements by $2.7$ standard deviations.

The present theoretical knowledge may be summarized as:
\begin{eqnarray}
\Delta\nu^{\rm th}&=&\Delta\nu^{LO}\left\{1
-\frac{\alpha}{\pi}\left(\frac{32}{21}+\frac{6}{7}\ln2\right)
-\frac{5}{14}\alpha^2\ln{\alpha}\right.
\nonumber\\
&+&\left(\frac{\alpha}{\pi}\right)^2
\left[\frac{1367}{378}-\frac{5197}{2016}\pi^2
+\left(\frac{6}{7}+\frac{221}{84}\pi^2\right)\ln2\right.
\nonumber\\
&-&\left.\frac{159}{56}\zeta(3)\right]+\left(\frac{\alpha}{\pi}\right)^3\left[
-\frac{3}{2}\pi^2\ln^2{\alpha}
+\left(-\frac{62}{15}\right.\right.
\nonumber\\
&+&\left.\left.\left.\frac{68}{7}\ln2\right)\pi^2\ln{\alpha}+D\right]\right\},
\label{eq::th}
\end{eqnarray}
where $\Delta\nu^{LO} ={7\over 12}\alpha^4 m_e$ is the leading-order result
\cite{Pirenne:1947,Berestetski:1949,Ferrell:1951zz}. The first-order term in
Eq.~(\ref{eq::th}) has been computed in Ref.~\cite{Karplus:1952wp}. The
second-order corrections have been derived by several
authors~\cite{Brodsky:1966vn,Barbieri:1972as,Bodwin:1977ut,Caswell:1978vz,Sapirstein:1983xr,
Adkins:1988nd,Adkins:1993zz,Hoang:1997ki,Pachucki:1997vm,Pachucki:1997zz,
Czarnecki:1998zv,Adkins:1998zz,Burichenko:2000qp}. In the order
$\alpha^7m_e$ the double-logarithmic \cite{Karshenboim:1993} and the
single-logarithmic terms \cite{Hill:2000zy,Melnikov:2000zz,Kniehl:2000cx} are
known, while the nonlogarithmic coefficient $D$ is not yet available. Including
all the terms known so far, we have \cite{Kniehl:2000cx}
\begin{equation}
\Delta\nu^{\rm th}=203.391\,69(41)\,\mbox{GHz},
\label{eq::thnum}
\end{equation}
where the error is estimated by the size of the third-order nonlogarithmic
contribution to the HFS in muonium atom \cite{Nio:1997fg}, which however does
not include annihilation and recoil effects. The result~(\ref{eq::thnum})  is
above the   experimental values~(\ref{eq::exp1}) and~(\ref{eq::exp2}) by $2.6$
and $3.5$ standard deviations, respectively. At the same time, it is only $1.2$
standard deviations below the most recent result~(\ref{eq::exp3}). Thus the
status of the QED prediction for positronium HFS remains ambiguous. 

Much activity is currently on the way to improve the experimental precision
\cite{Yamazaki:2012yr,Cassidy:2012}. On the theoretical side the accuracy is
limited by the unknown  third-order coefficient $D$. The corresponding
uncertainty is only two to four times smaller than the experimental one and soon
may become a limiting factor in the study of positronium HFS. On the other hand,
the calculation of the nonlogarithmic third-order term  in Eq.~(\ref{eq::thnum})
would result in one of the most precise theoretical predictions in physics.
This calculation, however, is an extremely  challenging problem of perturbative
quantum field theory complicated by the presence of  multiple scales and
bound-state dynamics.

In this Letter we make the first  major step towards the solution of this
problem and present the complete result for the  ${\cal O}(m_e\alpha^7)$
one-photon annihilation  contribution. The perturbative corrections to HFS split
into nonannihilation (radiative, radiative-recoil, and recoil corrections),
one- and multiple-photon annihilation contributions.  The nonannihilation and
one-photon annihilation  parts constitute about 47\% and 32\% of the
second-order nonlogarithmic correction, respectively. Thus the one-photon
annihilation contribution to the coefficient $D$ presumably gives a significant
fraction of the total nonlogarithmic third-order correction.

In the following, we briefly outline our method of calculation. Perturbation
theory of the positronium bound state has to be developed about the
nonrelativistic Coulomb approximation rather than free electron and positron
states. This can be done within the nonrelativistic effective field theory
\cite{Caswell:1985ui}, which is a systematic way to separate the multiple scales
characteristic to the bound-state problem. The bound-state dynamics involves
three different scales: the hard scale of electron mass $m_e$, the soft scale of
the bound-state three-momentum $\alpha^{} m_e$, and the bound-state energy
$\alpha^2 m_e$. Integrating out the hard and soft degrees of freedom results in
the potential nonrelativistic QED (pNRQED) \cite{Pineda:1997bj}, an effective
Schr\"odinger theory of a nonrelativistic electron-positron pair interacting
with ultrasoft photons, which is a relevant framework for the calculation of the
QED corrections to the positronium spectrum. We use  dimensional regularization
to deal with spurious divergences which appear in the process of scale
separation. Systematic use  of dimensional regularization
\cite{Czarnecki:1998zv,Pineda:1997ie,Kniehl:2002br} based on the asymptotic
expansion approach \cite{Beneke:1997zp,Smirnov:2002pj} is instrumental for the
high-order analysis as it provides ``built in'' matching of the effective theory
calculations to full QED.

\begin{figure}
\includegraphics[width=3.5cm]{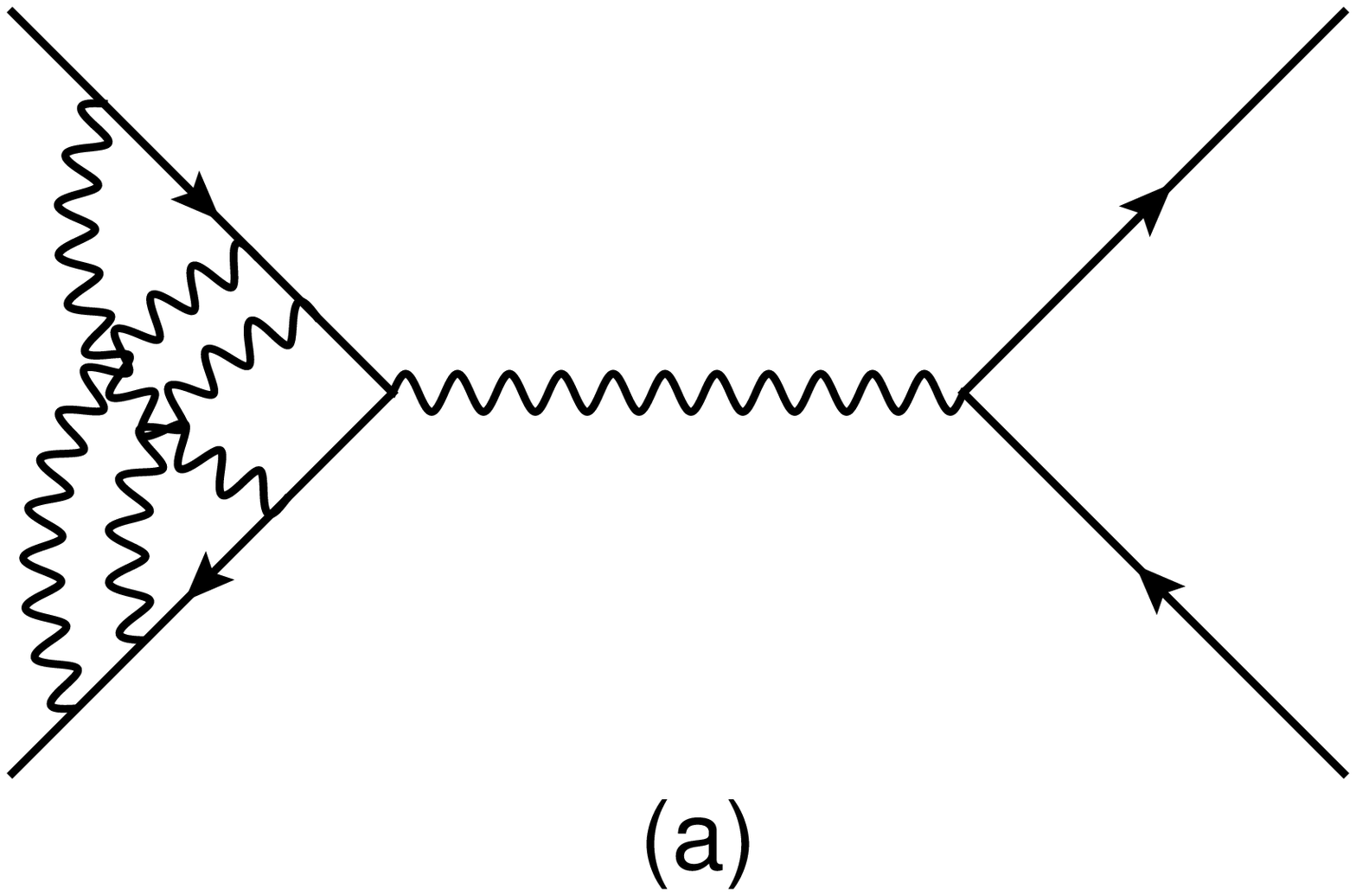}
\hspace*{5mm}\includegraphics[width=3.5cm]{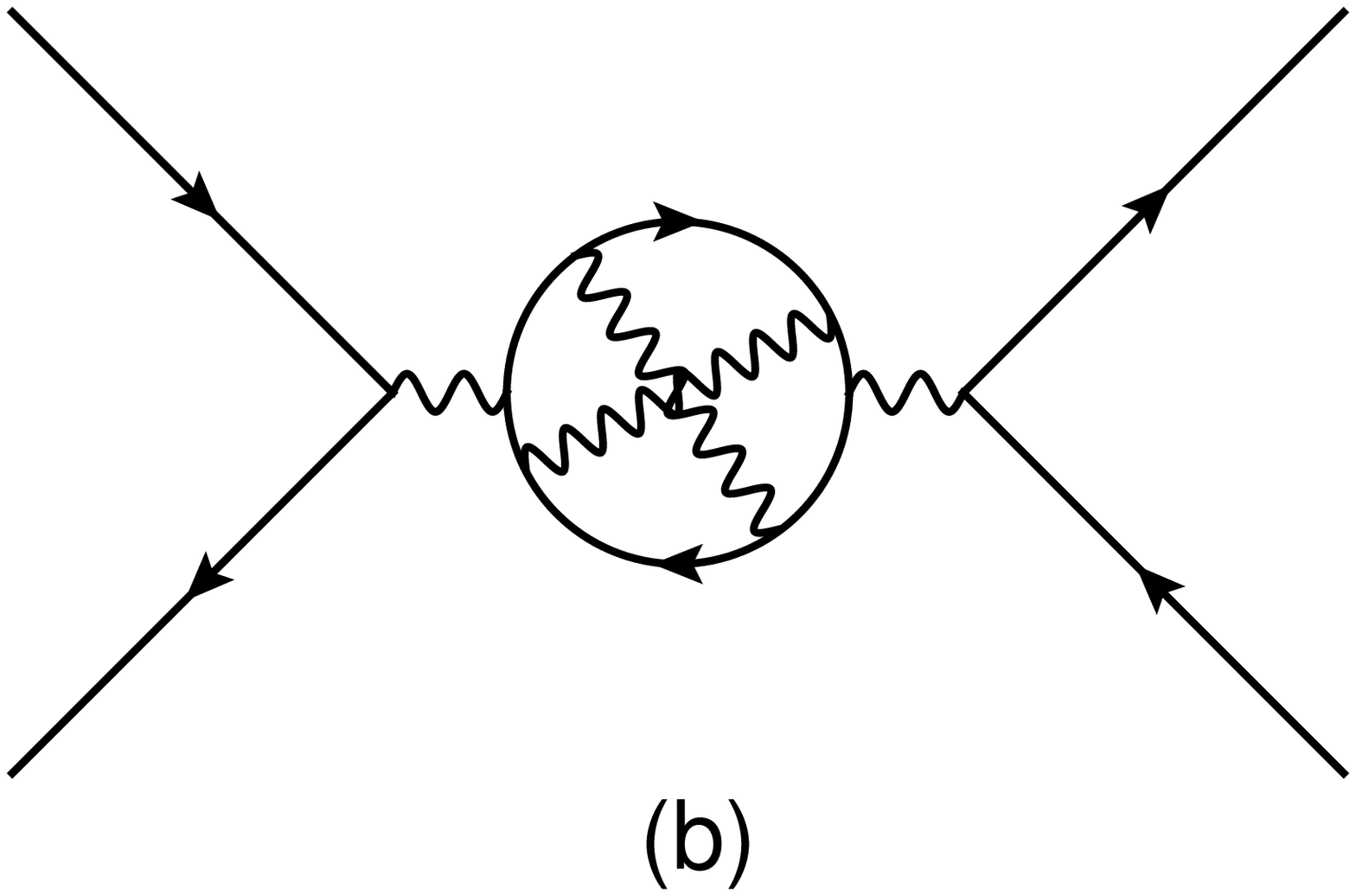}
\caption{\label{fig::diagrams} Three-loop Feynman diagrams contributing to (a) $R_o$
and (b) $P_o$.}
\end{figure}

The positronium HFS is given by the difference between the binding energy of the
ortho and parapositronium states  $\Delta\nu=E_o-E_p$. The leading order result
can be written as $\Delta\nu^{LO}=\left(\left[{1\over 3}\right]_{sct}
+\left[{1\over 4}\right]_{ann}\right)\alpha^4 m_e$, where nonannihilation
(scattering) and one-photon annihilation contributions are given separately. By
spin/parity conservation only the orthopositronium state is affected by the
one-photon annihilation. The corresponding correction to the binding energy
$E_o$ can be obtained by studying the threshold behavior of the vacuum
polarization function $\Pi(q^2)$
\begin{equation}
\left(q_\mu q_\nu-g_{\mu \nu}q^2\right)\Pi(q^2)=
i\int d^dx\,e^{iqx}\,\langle 0|Tj_{\mu}(x)j_{\nu}(0)|0\rangle.
\label{eq::vacpol}
\end{equation}
where $j_\mu$ is the electromagnetic current, $q^2=\left(2m_e+E\right)^2$ and
$E$ is the energy counted from the threshold. Only one-particle irreducible
contributions are retained on the right-hand side of Eq.~(\ref{eq::vacpol}) and
the on-shell renormalization of the QED coupling constant requires $\Pi(0)=0$.
The vacuum polarization function has bound-state poles at approximately Coulomb
energies $E_n^{C}=-\alpha^2m_e/(4n^2)$ with spin  (orbital) angular momentum
$S=1$ ($l=0$). Near the orthopositronium ground-state energy $E_o=E_1^C+{\cal
O}(\alpha^4)$ it reads
\begin{equation}
\raisebox{0pt}{$\rm lim$}
\raisebox{-6pt}{$\hspace*{-20pt}_{E\to E'_o}$}
\Pi(q^2)= {\alpha\over 4\pi}{R_o\over E/E'_o-1-i\varepsilon},
\label{eq::vppole}
\end{equation}
where $E'_o$ stands for $E_o$ without the total one-photon annihilation
contribution. The pole position differs from the physical orthopositronium mass
since the vacuum polarization function is defined as the one-particle
irreducible contribution to the current correlator~(\ref{eq::vacpol}). By
subtracting the pole one gets the regular part of the  vacuum polarization
function at $E=E'_o$
\begin{equation}
P_o=\raisebox{0pt}{$\rm lim$}
\raisebox{-6pt}{$\hspace*{-20pt}_{E\to E'_o}$}\left(e^2\Pi(q^2)
-{\alpha^2R_o\over E/E'_o-1-i\varepsilon}\right).
\label{eq::vpreg}
\end{equation}
Within the quantum-mechanical perturbation theory of pNRQED it is
straightforward to derive the following expression for the one-photon
annihilation contribution to the HFS
\begin{equation}
\Delta^{1-\gamma}_{ann}\nu=\Delta^{1-\gamma}_{ann} E_o
={\alpha^4 m_e\over 4}{R_o\over 1+P_o}.
\label{eq::master}
\end{equation}
The factor $R_o$ in this equation has a natural interpretation: annihilation is
a local process which probes the positronium wave function at the origin and the
residue of Eq.~(\ref{eq::vppole}) defines this quantity  in full QED beyond
nonrelativistic quantum mechanics.  On the other hand the factor  $1/(1+P_o)$
results  from the Dyson resummation of the vacuum polarization corrections  to
the off-shell  photon propagator in the annihilation amplitude.
Eq.~(\ref{eq::master}) can be computed order by order in perturbation theory
\begin{equation}
\Delta^{1-\gamma}_{ann}\nu={\alpha^4 m_e\over 4}\left[1+
\sum_{n=1}\left({\alpha\over \pi}\right)^nh^{(n)}\right],
\label{eq::hfsser}
\end{equation}
where the coefficients $h^{(n)}$ are determined by the series
$R_o=1+\sum_{n=1}\left({\alpha\over \pi}\right)^nr^{(n)}$ and
$P_o=\sum_{n=1}\left({\alpha\over \pi}\right)^np^{(n)}$ so that
$h^{(1)}=r^{(1)}-p^{(1)}$ and so on. For the calculation of the third order
corrections to the HFS we need all coefficients $r^{(n)}$ and $p^{(n)}$ up to
$n=3$. Typical three-loop Feynman diagrams contributing to  $R_o$ and  $P_o$ are
presented in Fig.~\ref{fig::diagrams}.

The first-order  coefficients    get only a one-loop hard contribution
$r^{(1)}=-4$ and $p^{(1)}=8/9$, which yields  $h^{(1)}=-{44/9}$. In the second
order the soft scale starts to contribute and one has to take into account an
arbitrary number of Coulomb photon exchanges. The second-order correction to
$R_o$ can be read off the QCD result for the photon-mediated heavy quarkonium
production rate
\cite{Melnikov:1998ug,Penin:1998kx,Czarnecki:1997vz,Beneke:1997jm} by adopting
the QED group
factors $C_F=1$, $C_A=0$, $T_F=1$, as well as the number of the light (heavy)
fermions $n_l=0$ ($n_h=1$)
\begin{equation}
r^{(2)}={527\over 36}+\left(-{2\over 3}\ln{\alpha}-{235\over 72}
+2\ln{2}\right)\pi^2-\zeta(3),
\label{eq::r2}
\end{equation}
where $\zeta(3)=1.20206\ldots$ is a value of Riemann's zeta-function. By
using the method outlined above we evaluate the  second-order correction to
$P_o$ with the result
\begin{equation}
p^{(2)}={3\over4}+\left(-\ln{\alpha}+{27\over 16}-{\ln{2}\over 2}\right){\pi^2\over 2}
-{21\over 8}\zeta(3).
\label{eq::pip2}
\end{equation}
This gives
\begin{equation}
h^{(2)}={1477\over 81}+
\left(-{\ln{\alpha}\over 6}-{1183\over 288}+{9\over 4}\ln{2}\right)\pi^2
+{13\over 8}\zeta(3),
\label{eq::h2}
\end{equation}
in agreement with Ref.~\cite{Hoang:1997ki}.

The third-order coefficients get contributions from all the scales present in
the problem. By  adjusting the QCD results \cite{Beneke:2007pj,Beneke:2007gj} we
obtain the following expression
\begin{eqnarray}
r^{(3)}&=&
-{383\over 18}+ \left[-{3\over 2}\ln^2{\alpha}
+\left(-{7\over 90}+8\ln{2}\right)\ln{\alpha}\right.
\nonumber\\
&-&\left.{1019\over 180}-4\ln{2}+\delta_o^{us}\right]\pi^2
+2\zeta(3)-{109\over 864}\pi^4 +2c_{v\,0}^{(3)}.
\nonumber\\
&&
\label{eq::r3}
\end{eqnarray}
Here $\delta_o^{us}$ is an analog of the Bethe logarithm in hydrogen Lamb shift,
which parametrizes the ultrasoft contribution \cite{Beneke:2007pj}.  It does
not scale with the group factors and requires independent evaluation in the QED
case, which gives $\delta_o^{us}=18.8646(17)$.  The coefficient $c_{v\,0}^{(3)}$
in Eq.~(\ref{eq::r3})  parametrizes the third-order hard contribution to the
Wilson coefficient in the effective theory decomposition of the vector current
${\bfm j}=c_v\psi^\dagger{\bfm \sigma}\chi+\ldots$  in terms of the
nonrelativistic electron  and positron two-component Pauli spinors  $\psi$ and
$\chi$. The third-order term of the  perturbative series
$c_v=1+\sum_{n=1}^\infty({\alpha\over\pi})^nc^{(n)}_v$ is given by the
three-loop vertex diagrams (see {\it e.g.} Fig.~\ref{fig::diagrams}(a))
evaluated at the threshold  and has been recently computed in QCD
\cite{Marquard:2014pea}. The coefficients of the series  are in general infrared
divergent.  These spurious divergences result from the scale separation in the
effective theory framework and cancel out in the final result for physical
observables. The  value $c_{v\,0}^{(3)}=35.76 \pm 0.53$ corresponds to the
coefficient $c^{(3)}_v$ defined within the $\overline{\rm MS}$ subtraction
scheme at the renormalization scale $\mu=m_e$. The logarithmic part of
Eq.~(\ref{eq::r3}) agrees with Ref.~\cite{Kniehl:2002yv}. The third-order term
in Eq.~(\ref{eq::vpreg}) reads
\begin{equation}
p^{(3)}=\left(2\ln{\alpha}-3\right)\pi^2+p^{(3)}_{h\,0},
\label{eq::pip3}
\end{equation}
where the last term parametrizes the third-order hard contribution  given by the
three-loop vacuum polarization diagrams (see {\it e.g.}
Fig.~\ref{fig::diagrams}(b)) evaluated at the threshold. As in the case of the
vertex correction,  this quantity is infrared divergent and the  coefficient
$p^{(3)}_{h\,0}=0.16\pm 0.04$ corresponds to the $\overline{\rm MS}$ subtraction
scheme with $\mu=m_e$. By adding up all the relevant terms we get
\begin{equation}
h^{(3)}=-{3\over 2}\pi^2\ln^2{\alpha}+\left(-{1181\over 270}
+8\ln{2}\right)\pi^2\ln{\alpha}+h^{(3)}_0,
\label{eq::h3}
\end{equation}
where the nonlogarithmic part reads
\begin{eqnarray}
h^{(3)}_0&=&-{49309\over 1458} +\left({16573\over 3240}-{65\over 9}\ln{2}+
\delta_o^{us}\right)\pi^2-{221\over 18}\zeta(3)
\nonumber \\
&&-{109\over 864}\pi^4
+2c^{(3)}_{v\,0}-p^{(3)}_{h\,0},
\label{eq::h30}
\end{eqnarray}
or numerically $h^{(3)}_0=197.8 \pm 1.1$. From the effective theory point of
view the structure of the third-order logarithmic corrections in the one-photon
annihilation  contribution to the positromium HFS is identical to the
orthopositronium three-photon decay width.  The coefficients of the logarithmic
terms in Eq.~(\ref{eq::h3}) do agree with the series for the  width
\cite{Kniehl:2000dh,Hill:2000qi,Melnikov:2000fi}  up to a substitution
of the coefficient ${A_o\over 3}\to -{2\over 3 }h^{(1)}$ in the
interference term between the one-loop and the two-loop single-logarithmic
corrections.

Finally for the third-order nonlogarithmic  one-photon annihilation contribution
to the HFS we obtain
\begin{equation}
D_{ann}^{1-\gamma}= {3\over 7}h^{(3)}_0=
84.8 \pm  0.5\,, \label{eq::Dann}
\end{equation}
which is the main result of this Letter. The coefficients of the third-order
corrections to HFS in positronium and muonium atom \cite{Nio:1997fg} are
compared in Table~\ref{tab::posvsmuon}. It is interesting to note that the
ultrasoft contribution due to $\delta_o^{us}$ approximates the complete
result~(\ref{eq::Dann}) with 5\% accuracy. The nonannihilation contribution
includes a similar term and we may speculate that it is also dominated by the
ultrasoft contribution. This does not seem implausible since the fully
relativistic corrections from the hard scale are known to usually be suppressed.
For example, the pure radiative corrections to the HFS related to the electron
anomalous magnetic moment $a_e$,
$\Delta_{a_e}\nu=(\alpha^4m_e/4)\left[(1+a_e)^2-1\right]$, gives only  a tiny
contribution $D_{a_e}=1.16229\ldots$, where we used the two and three-loop
result for $a_e$ \cite{Barbieri:1972as,Laporta:1996mq}.  In this case the
nonannihilation contribution would be given by  $D_{sct}\approx {4\pi^2\over 7}
\delta_o^{us}\approx 106$, which slightly exceeds  the one-photon annihilation
contribution~(\ref{eq::Dann}) in full analogy with the structure of the
second-order corrections. Then we get an estimate $D\approx 191$, which is close
to the muonium result.
\begin{table}[t]
  \begin{ruledtabular}
    \begin{tabular}{c|c|c|c}
      & $\ln^2{\alpha}$ & $\ln{\alpha}$ & $D/\pi^2$\\
      \hline
      Positronium &  $-{3\over 2}$ & $-\frac{62}{15}+\frac{68}{7}\ln2\approx 2.6001$ &
      $8.59(5)^{1-\gamma}_{ann}$\\
      Muonium & $-{8\over 3}$ & $-{281\over 180}+{8\over 3}\ln2\approx 0.2873$ & $16.233$ \\
    \end{tabular}
    \end{ruledtabular}
    \caption{\label{tab::posvsmuon}
      The   coefficients of $\alpha^3/\pi$ in perturbative series for positronium and muonium HFS.}
\end{table}

To summarize, in this Letter we presented the ${\cal O}(\alpha^7m_e)$ one-photon
annihilation contribution to the positronium HFS, which is  the first nontrivial
third-order QED result in positronium spectroscopy  beyond the logarithmic
approximation. This opens a prospect of advancing the theoretical analysis of
positronium to a completely different level of precision. Our final prediction
for the positronium HFS including  the ${\cal O}(\alpha^7m_e)$ one-photon
annihilation term reads
\begin{equation}
\Delta\nu^{\rm th}=203.391\,91(22)\,\mbox{GHz}\,.
\label{eq::final}
\end{equation}
The error due to the missing part of the  ${\cal O}(\alpha^7m_e)$ corrections is
estimated by the size of the evaluated one-photon annihilation
contribution~(\ref{eq::Dann}) and is reduced by a factor of two with respect to
the previous estimate based on the  size of the nonlogarithmic corrections to
the muonium HFS. This agrees with an error estimate based  on the approximation
of the missing nonannihilation correction by the ultrasoft contribution
discussed above. At the same time if we include the  nonannihilation correction
approximated in this way into the numerical analysis, our central value  changes
to  $\Delta\nu^{\rm th}=203.392\,11\,\mbox{GHz}$, {\it i.e.} gets within one
standard deviation from the most recent  experimental result~(\ref{eq::exp3}).

\begin{acknowledgements}
This work has been supported by the DFG SFB/TR~9 ``Com\-pu\-ter\-gest\"utzte
Theoretische Teil\-chen\-physik''. P.M was supported in part by the EU Network
{\sf LHCPHENOnet} PITN-GA-2010-264564 and {\sf HIGGSTOOLS} PITN-GA-2012-316704.
The work of A.P. was supported in part by NSERC,  Alberta Ingenuity Foundation,
and Mercator DFG grant.
\end{acknowledgements}


\end{document}